\documentclass[%
 aip,
 amsmath,amssymb,
 reprint,%
]{revtex4-1}

\usepackage{graphicx}
\usepackage{dcolumn}
\usepackage{bm}

\usepackage[utf8]{inputenc}
\usepackage[T1]{fontenc}
\usepackage{mathptmx}
\graphicspath{{figures/}}

\begin{document}

\preprint{AIP/123-QED}

\title[Diffusion regions and 3D energy mode development in spontaneous reconnection]{Diffusion regions and 3D energy mode development in spontaneous reconnection}

\author{Shuoyang Wang}%
\affiliation{Department of Earth and Planetary Science, The University of Tokyo, Bunkyo-ku, Tokyo 113-0033, JAPAN}%

\author{Takaaki Yokoyama}
\affiliation{Department of Earth and Planetary Science, The University of Tokyo, Bunkyo-ku, Tokyo 113-0033, JAPAN}%

\date{\today}

\begin{abstract}
The understanding of magnetic reconnection in three-dimension (3D) is far
shallower than its counterpart in two-dimension (2D) due to its potential
complication, not to mention the evolving of the spontaneously growing 
turbulence. 
We investigate the reason for reconnection 
acceleration on the characters and development of diffusion regions 
and sheared 3D energy modes (energy modes that are not parallel to the 
anti-parallel magnetic fields) during the turbulence building stage.
We found that multiple reconnection layers emerge due to the growth 
of 3D sheared tearing instability. Diffusion regions on adjacent reconnection 
layers form an inflow-outflow coupling that enhances the local reconnection. 
Further coupling of the existing energy modes breeds new energy modes near 
the current sheet edge. As reconnection layers span and interact with each 
other across the whole current sheet, global magnetic energy consumption 
accelerates. The significant contribution of 3D energy modes and their 
interaction to the reconnection rate enhancement seems to be independent 
on magnetic diffusivity. On the other hand, the global guide field 
changes the layout of 3D reconnection layer thus determines whether the 
system is fast-reconnection-preferable.
\end{abstract}

\maketitle

\section{\label{sec:level1}Introduction}

Magnetic reconnection is one of basic nonetheless complicated plasma 
processes. A reconnecting extensive current sheet spontaneously generates 
turbulence. The potential nonlinear nature of reconnection frequently tangles 
with turbulence in both macro \citep{kow09,ois15,hua16,kow17,ber17} and micro 
scales \citep{nak16,lou17a2}.

The understanding of the interaction between turbulence and reconnection 
starts from how turbulence changes the reconnection rate. 
It was found in two-dimensional (2D) simulation that the global reconnection 
rate increases by introducing a strong random perturbation onto a current sheet 
\citep{mat85}. It was also shown in three-dimensional (3D) simulation that 
driven turbulence increases the global reconnection rate which becomes 
independent of diffusivity \citep{kow09}. 
The energy of turbulence is applied to the system manually in both cases. 
How the self-generated turbulence accelerates the reconnection in a free system 
is not sufficiently understood.

One turbulence reconnection model theoretically describes the reconnection 
development in a 3D current sheet when a statistical steady turbulent state 
has reached \citep{laz99}. In the global scale, which corresponds to the energy 
input scale, the magnetic fields are absorbed into the global current sheet 
as the turbulent area expands across the current sheet. In the local scale, 
which corresponds to the dissipation scale, reconnection happens in the 
diffusion region and blends the field lines. A fast global reconnection rate is 
deduced by coupling to the MHD turbulence theory \citep{gol95} that the local 
reconnection in local diffusion regions is fast. 

Fast overall reconnection is found in 3D spontaneous reconnection. In a current 
sheet with no guide field, the reconnection rate is shown to be in the magnitude 
of $0.001$, while a non-monotonic change of reconnection rate with diffusivity is 
found \citep{ois15}. In a current sheet with a guide field that varies from half to 
twice of the strength of the anti-parallel fields, the overall reconnection rate is 
consistently high that $M_A \sim 0.01$. This result is found to be independent of 
diffusivity. The reconnection rate remains high even when the guide field is 
reduced to $10\%$ of the anti-parallel field \citep{kow17}. Judging by the 
up-to-date results, it seems that 3D reconnection has the potential to be fast 
in the system with a large Lundquist number. On the other hand, physics might 
change with the global guide field strength.

The global features, such as the fast global reconnection rate and the global 
current sheet expansion \citep{ois15,hua16,ber17,kow17}, are often discussed. 
The validation of the local reconnection process predicted by the turbulence 
theory has not been confirmed  due to the structural complexity. 
In addition, the present MHD turbulence theories all consider a strong background 
field \citep[e.g.,][]{gol95,bol17,mal17}. It limits the application of the present 
turbulence reconnection model. Thus detailed descriptions of both local 
reconnection in diffusion regions and global behavior are needed to explain more 
general reconnection picture.

It was speculated that multiple tearing layers might emerge if a multi-modal 
perturbation is applied to a 3D large sheared magnetic field \citep{gal77}.
In 3D numerical simulations, multiple reconnection layers coexisting 
in a current sheet built by magnetic shear in large scale are 
frequently observed \citep{ono04,lan08,wang15,hua16,ber17,kow17}. 
If these reconnection layers are independent of each other, each 
layer might follow the scenario of 2D plasmoid instability 
\citep{shi01,lou05,hua10}. 
As a whole, they contribute to a global fast reconnection. 
If the distance between layers is small than their width, these 
layers might interact with each other. 
As the interaction is along the sheet width direction (the magnetic 
shear direction), it should change the plasma and magnetic fields 
going into the diffusion regions on each layer.
In this paper, we execute numerical simulations to understand the 
development of a spontaneous reconnecting current sheet when multiple
reconnection layers emerge. How diffusion regions are affected by 
the interaction is the main topic.

From our previous study, we noticed an inflow-outflow coupling of flow 
in a global current sheet with multiple reconnection layers \citep{wang15}. 
The coupling is observed as a direct feeding of plasma flow from 
the outflow region of a diffusion region on one reconnection layer to 
the inflow region of a diffusion region on another reconnection layer.
However, as the initial perturbation was a random seed of velocity 
field, it is difficult to extract a detailed analysis of the local 
diffusion region. 

In order to get a clear picture of how diffusion regions on different 
reconnection layers interact with each other,  we apply only 2 
reconnection layers in the current sheet from the beginning. 
This can be achieved by perturbing the current sheet with 
selected tearing instability eigenfunctions. 
The inflow-outflow coupling is expected to be built once the 
pair of reconnection layers are close to each other.
By scrutinizing the details of reconnection layer coupling, the physics 
in the local and global development of diffusion regions is thought to 
be revealed. The result is believed to compensate for the insufficient 
detailed description in turbulence reconnection theory 
\citep{laz99,ber17}.

The configuration of this paper is as the following. 
In Sec.\ref{sec:model}, we introduce the simulation model and setup. 
In Sec.\ref{sec:result1}, the local and global detailed analyses of 
a typical tearing-eigenfunction-perturbed simulation are presented, 
followed by the result of a parameter survey of changing tearing 
mode wavelength. 
In Sec.\ref{sec:result2}, the result found in the previous section 
is reexamined in random-perturbed systems with various diffusivities 
and strengths of the global guide field. 
In Sec.\ref{sec:discussion}, discussion on the extension of our 
present result is shown.
Final conclusion is summarized in Sec.\ref{sec:conclusion}.

\section{Simulation model}\label{sec:model}

We apply ordinary one-fluid resistive (uniform diffusivity) MHD equations 
which neglect viscosity, gravity and heat conduction for simplicity, 
together with the equation of state for the ideal gas:
\begin{eqnarray}
  \label{eq:masseqset}
  \frac{\partial \rho}{\partial t}+({\bf v} \cdot \nabla) \rho
     & = & -\rho (\nabla \cdot {\bf v}) \\
  \label{eq:momentumeqset}
  \rho \frac{\partial {\bf v}}{\partial t}
           + \rho ({\bf v} \cdot \nabla ){\bf v}
     & = & -\nabla p+\frac{{\bf J}\times {\bf B}}{c} \\
  \label{eq:energyeqset}
  \rho \frac{\partial e}{\partial t}+\rho \left( {\bf v} \cdot \nabla \right) e 
     & = & - p \nabla \cdot {\bf v} + \eta {\bf J}^2 \\
  \label{eq:inducteqset}
  \frac{\partial {\bf B}}{\partial t}
     & = & \nabla \times \left( {\bf v} \times {\bf B} - c \eta {\bf J} \right) \\
  \label{eq:ampereeqset}
  {\bf J}
     & = & \frac{c}{4\pi }\nabla \times {\bf B}
\end{eqnarray}
in which $\rho$ and $p$ are plasma mass density and pressure, 
$\gamma = 5/3$ is the adiabatic index, 
$e$ is the internal energy per unit mass that 
$e = p/[(\gamma-1)\rho]$ and $\eta$ is the resistivity. 

All quantities are normalized by the typical parameters. 
The length scale is in the unit of the initial current sheet 
width $\delta$. The time scale is normalized by 
$t_{A} = \delta/v_{A0}$, where $v_{A0}$ is the asymptotic 
Alfv$\acute{\text{e}}$n velocity. Initial uniform mass density 
$\rho_{0}$ is used for normalizing mass density. 
Normalization of the magnetic field is implemented by 
$B_{0}=v_{A0}\sqrt{\rho_{0}}$. Current density is normalized by 
$J_{0} = c B_{0}/\delta$ while plasma pressure is normalized by 
$p_{0} = B_{0}^{2}$. All equations are solved in Cartesian coordinate. 

Our background magnetic field is composed of anti-parallel component 
$B_y(x)$ with a uniform finite guide field $B_z$:
\begin{eqnarray} \label{eq:Binitial}
  {\bf B} & = & B_{y}{\bf \hat{y}}+B_{z}{\bf \hat{z}} \nonumber \\
          & = & B_{y0} \tanh \left( \frac{x}{a} \right) 
          \left\{ \frac{1}{2} \bigg[ \tanh \left( 
          \frac{\mid x \mid - 4\delta}{a} \right) 
          -1 \bigg] \right\} {\bf \hat{y}} \nonumber \\
          & + & \alpha B_{y0} {\bf \hat{z}}
\end{eqnarray}
where $B_{y0} = \sqrt{4\pi}B_{0}$ and $a = 0.5\delta$. 
This modified Harris sheet has one central global current 
sheet centered at $x = 0$ and two secondary current sheets near 
$x = \pm 4\delta$.
This configuration is to simplify the boundary condition across 
the current sheet direction, while the secondary current sheets do 
not participate in reconnection essentially within our simulation 
time.
The guide field strength is mediated by parameter $\alpha$.
To obtain a static global structure, the total uniform pressure 
is set as:
\begin{equation} \label{eq:ptotal}
  P_{\text{tot}} = \frac{B_0^2}{2} (1+\alpha^2) (1+\beta)
\end{equation}
in which $\beta = 0.2$ is the ratio between the plasma pressure 
and the magnetic pressure at the asymptotic field outside of the 
central global current sheet. 

We divide the simulations into two groups regarding the different 
forms of the perturbation onto the initially static central global 
current sheet. Other parameters are also changed for a more general 
scan.

We denote the first simulation group as the eigenmode-perturbation 
simulation group. The simulations in this group are triggered by 
velocity and magnetic fields of selected tearing mode eigenfunctions. 
They are used to understand the basic development of the reconnection 
layer interaction.
In this group, two subgroups of simulations are included 
to make comparisons. 

The first subgroup, called double-layer simulation, is perturbed by two 
superposed tearing modes ${\bf k_L}$ and ${\bf k_R}$. These two modes 
are 3D tearing modes which are shear to the global anti-parallel 
magnetic field. The perturbation results in the emergence of one 
reconnection layer on either side of the central global current sheet 
from the beginning of the simulation (similar to 
\citet{gra07}). 

To simplify the analysis, the two modes ${\bf k_L}$ and ${\bf k_R}$ 
are chosen to be rotational-symmetric across the central global current 
sheet center.
The initial fields become 
\begin{eqnarray} 
  \label{eq:Btotdouble}
  {\bf B}(t = 0) & = & {\bf B^{[0]}} + {\bf B^{[1]}_L} + {\bf B^{[1]}_R} \\
  \label{eq:Vtotdouble}
  {\bf v}(t = 0) & = & {\bf v^{[1]}_L} + {\bf v^{[1]}_R},
\end{eqnarray}
where $[0]$ and $[1]$ represent background and first-order perturbation 
components of fields.
The amplitude of the initial perturbation is determined that
\begin{equation} \label{eq:v0norm}
  \text{max}(|v_x|,|v_y|,|v_z|) = 0.01v_{A0}. 
\end{equation}
This amplitude is small enough to start from the linear stage but 
large compared to many other studies. Since what we are interested in 
is the nonlinear interaction between the two reconnection layers, 
by applying appropriately large initial amplitude, the system enters 
the nonlinear stage earlier that the simulation resource is saved.

The other subgroup, called single-layer simulation, is perturbed by 
one single 3D tearing mode ${\bf k_R}$ that only one reconnection 
layer emerges on the positive-$x$ side of the central global current 
sheet. The initial fields are 
\begin{eqnarray} 
  \label{eq:Btotsingle}
  {\bf B}(t = 0) & = & {\bf B^{[0]}} + {\bf B^{[1]}_R} \\
  \label{eq:Vtotsingle}
  {\bf v}(t = 0) & = & {\bf v^{[1]}_R}.
\end{eqnarray}
This group of simulations is used to differentiate the reconnection 
layer interaction result from the potential nonlinear development of 
a single tearing mode.

We label each tearing mode as $(m, n)$. The integers $m$ and $n$ are 
defined by the box size $L_y$ and $L_z$ along $y$- and $z-$direction 
respectively that 
\begin{eqnarray}
  \label{eq:mdef}
  m & = & k_y \frac{L_y}{2\pi} \\
  \label{eq:ndef}
  n & = & k_z \frac{L_z}{2\pi},
\end{eqnarray}
where $k_y$ is always positive while $k_z$ can be positive or negative. 
In our system, the mode growing on the negative-$x$ side is labeled as $n < 0$ 
while the mode growing on the positive-$x$-side is labeled as $n > 0$. 
Therefore, double-layer simulations are labeled in the form of $(m,\pm |n|)$ 
while single-layer simulations are labeled as $(m,|n|)$.
As a tearing mode $(m,n)$ grows, its correspondent energy mode 
$(m,n)$ also grows.

The tearing instability grows on the layer where
\begin{equation} \label{eq:kdB}
  {\bf B} \cdot {\bf k} = 0
\end{equation}
is satisfied. 
As the tearing instability grows, the layer gradually 
grows into a reconnection layer.
We define a resonance factor $q$ that 
\begin{equation} \label{eq:resonfack}
  q = \frac{B_y(x)}{B_z},
\end{equation}
which mimics the safety factor used in a tokamak or a 
reversed-field pinch (RFP). 
On a reconnection layer with a certain $q$, a series of 
harmonics grows with the same ratio of $n/m$ that
\begin{equation} \label{eq:resonfraction}
  q = -\frac{L_y}{L_z} \frac{n}{m}.
\end{equation}
By combining Eq.(\ref{eq:resonfack}) and Eq.(\ref{eq:resonfraction}), 
the resonance condition determines the reconnection layer 
position $x_s$ along the $x$-direction in the initial central 
sheared magnetic fields:
\begin{equation} \label{eq:layerpos}
  x_s \sim \alpha a \frac{L_y}{L_z} \frac{n}{m}
\end{equation}
for a certain tearing mode $(m,n)$ and its correspondent 
energy mode $(m,n)$.

The tearing modes we apply as ${\bf k_R}$ are $(1,1)$, $(2,1)$, $(3,1)$,
$(4,1)$ for the two subgroups. Their rotational-symmetric counterparts 
are also applied as ${\bf k_L}$ in the double-layer simulation subgroup.
Among them, we select the tearing mode $(3,1)$ as the typical mode to 
investigate the reconnection layer interaction across the sheet.
In all these simulations, the guide field strength $\alpha = 0.1$ and 
the diffusivity 
$\widetilde{\eta} = \eta c^2 / (4\pi) \sim 3.2\times10^{-4} \delta^2/t_A$ 
are used.
The double-layer simulation $(3,\pm 1)$ continues until the boundary 
is influencing the central global current sheet ($\sim 640t_A$). 
The other simulations stop at $\sim 200t_A$.
All single-layer simulations maintain a translational-invariance along 
${\bf k_R}$ until the end of simulations.

The second simulation group is denoted as the random-perturbation 
simulation group. It is used to confirm the result achieved from the
eigenmode-perturbation simulations.
The simulations in this group are triggered by a random
velocity field with an amplitude the same as Eq.(\ref{eq:v0norm}). 
We separate the simulations into 
two subgroups. In one subgroup, we vary the guide field strength $\alpha$ 
from $0.01$ to $0.2$, while the diffusivity is kept the same as the 
eigenmode-perturbation simulation group 
($\widetilde{\eta} = 3.2\times10^{-4}\delta^2/t_A$). 
In another subgroup, we test three cases of simulation with different 
diffusivity $\widetilde{\eta} = 1.6$, $3.2$ and $6.5\times10^{-4}\delta^2/t_A$, 
while the guide field is the same as the eigenmode-perturbation simulation 
group ($\alpha = 0.1$). 
The Lundquist number $S$ in our system is defined by the total 
Alfv$\acute{\text{e}}$n speed $v_{A} = v_{A0}\sqrt{1+\alpha^{2}}$ that 
$S = v_{A} L/\tilde{\eta} \sim 0.3$, $0.7$ and $1.3 \times 10^{5}$, 
respectively for the subgroup with different diffusivity.
All simulations in this group continue until the current sheets at the 
boundary impact strongly to the central global current sheet 
(roughly $700t_A \sim 900t_A$). 
We stop the simulation of $\alpha = 0.01$ case in the random-perturbation 
simulation group at $t = 1000t_A$ as we are interested in the 
global current sheet development in the same time range.

In all simulations, the simulation box has a periodic boundary condition 
on all sides. The box size is 
$L_x \times L_y \times L_z = 10\delta \times 24\delta \times 6\delta$, 
which is resolved by $640 \times 1000 \times 250$ grids at least. 
Across the $x$-direction, we apply non-uniform grids to resolve the 
central global current sheet that $\Delta x \geqslant 0.005\delta$. 
Uniform grids along $y$- and $z$-direction are adjusted with 
$\Delta y = \Delta z = 0.024\delta$. 
In double-layer simulation $(4,\pm 1)$ and its single-layer counterpart, 
$\Delta x \geqslant 0.004\delta$ and $\Delta y = \Delta z = 0.015\delta$ 
are used. For the higher Lundquist number case, 
$\Delta x \geqslant 0.0025\delta$ and $\Delta y = \Delta z = 0.015\delta$ 
are used. Resolution check is done by applying different widths of current 
sheet in solving 1D diffusion equation. The grid size we apply here 
could resolve the diffusion region with a thickness of at least 
$a = 0.02\delta$. We test the convergence check by changing the grid size 
into $\Delta y = \Delta z = 0.04\delta$ for the double-layer simulation 
$(1,\pm 1)$. At $t = 100t_A$, the reconnection rate increases by $3\%$. 
The typical structures are not essentially changed, only the local 
reconnection is reduced moderately in the higher resolution simulation. 
So we argue that the resolution does not have a strong impact on our 
model, however the scaling might be changed when an even higher 
resolution is applied. We use CIP-MOCCT code \citep{kud99} with artificial 
Lapidus-type viscosity\citep{lap67} developed by H. Isobe from Kyoto 
University.

\section{Simulation result I: Eigenmode-perturbation group}\label{sec:result1}
  \subsection{Typical case: double-layer simulation $(3,\pm 1)$}\label{sec:result1:typical}

We specifically study the development of the double-layer simulation 
$(3,\pm 1)$ as a typical case. The reconnection layer of the 
tearing mode $(3,1)$ locates close to the current sheet center that 
it can build an efficient interaction with its counterpart across 
the central global current sheet. Moreover, the tearing mode $(3,1)$ 
is one of the most unstable 3D tearing modes in our system. 
It is expected to be dominant in the linear phase.

    \subsubsection{Local analysis of inflow-outflow coupling}\label{sec:result1:typical:local}

\begin{figure}
   \centering
   \includegraphics[scale=0.25]{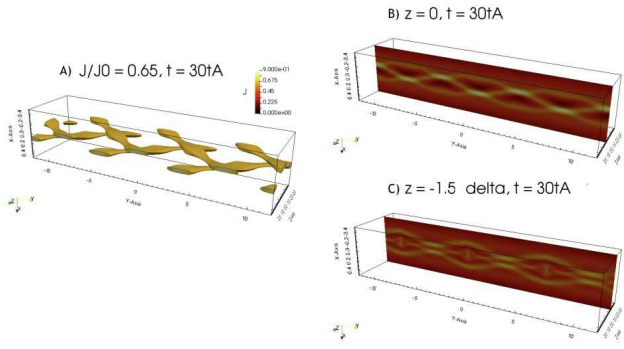}
   \caption{3D structures of the double-layer simulation 
            $(3,\pm 1)$ at $t = 30t_A$. 
            Bright features represent diffusion regions. 
            Panel A): current density isosurface $J/J_0 = 0.65$. 
            Panel B): current density $J$ on $z = 0$ plane.
            Panel C): current density $J$ on 
            $z = -1.5\delta$ plane.}
   \label{fig:3Ddouble03014}
\end{figure}

After simulation starts, two reconnection layers emerge. 
The isosurface of the current density $J$ is 
plotted in Fig.\ref{fig:3Ddouble03014} panel A). 
The diffusion regions are defined as the regions with high 
current density, which are the bright features in 
Fig.\ref{fig:3Ddouble03014}. 
Meanwhile, converging inflows and diverging outflows 
are identified in the enhanced current density region.
The diffusion regions form lines along the sheet direction. They are 
parallel to the local ${\bf B}$ and perpendicular to ${\bf k}$ of the
correspondent tearing mode. 
Since ${\bf k}$ on either side of the current sheet are different, 
the alignment of diffusion regions changes on different $z$-plane. 
Two characteristic alignment structures are shown in panels B) ($z = 0$) 
and C) ($z = -1.5\delta$). In panel B), tearing modes on either side has 
a phase shift of $\pi$ along the current sheet ($y$-direction), 
while in panel C) the phase shift is $\pm \pi/2$. We denote the planes 
similar to $z = 0$ and $z = -1.5\delta$ as anti-symmetric and symmetric 
planes, respectively. 

As can be seen in Fig.\ref{fig:3Ddouble03014}, the system varies 
from the anti-symmetric plane to the symmetric plane periodically if 
each $z$-plane is scanned along the global guide field direction. 
We assume that the regions in between the anti-symmetric and the 
symmetric planes are possessing the characters from both of these planes. 
So in the following part, we concentrate on the development of these two 
characteristic planes and treat the other planes as a mixture. 

\begin{figure}
   \centering
   \includegraphics[scale=0.36]{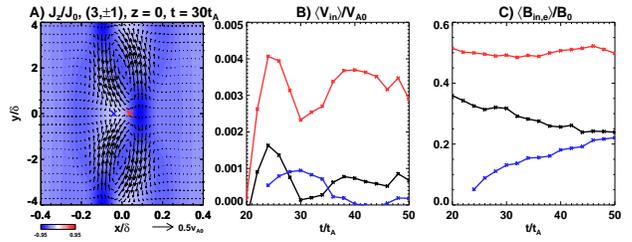}
   \caption{Panel A): the inflow-outflow coupling in the double-layer 
            simulation $(3,\pm 1)$ at $t = 30t_A$. 
            One wavelength is presented along the $y$-direction. 
            The velocity field on $z = 0$ plane is 
            plotted as vectors over the contour plot of current density 
            $J_z/J_0$. 
            The positions of plasma inflow and upstream magnetic field 
            measurement on $y = 0$ line are labeled by thick red and blue
            crosses. 
            Panel B): spatially averaged plasma inflow into the diffusion 
            region on the reconnection layer of the tearing mode $(3,1)$.
            Panel C): spatially averaged upstream magnetic field into 
            the diffusion region on the reconnection layer of the 
            tearing mode $(3,1)$. Red and blue lines with crosses are
            measurements on the anti-symmetric and symmetric planes in 
            the double-layer simulation $(3,\pm 1)$ respectively. 
            Black lines with crosses are measurements in the single-layer 
            simulation $(3,1)$. 
            The values are shown in the time range when diffusion regions 
            can be identified and maintain intact.}
   \label{fig:0301coupling2}
\end{figure}

The anti-symmetric plane is in favor of inflow-outflow coupling. 
A typical inflow-outflow coupling on $z = 0$ plane in the double-layer
simulation $(3,\pm 1)$ is shown as the velocity vectors in panel A) of 
Fig.\ref{fig:0301coupling2}. The diffusion regions are described as 
being coupled when the inflow-outflow coupling is built by them. 
To understand how the inflow-outflow coupling changes the local 
reconnection more quantitatively, 
we study the reconnection characters of a coupled diffusion region on 
an anti-symmetric plane (z = 0). The diffusion region is selected to 
be on the reconnection layer of the tearing mode $(3,1)$ (on the 
positive-$x$ side) and lays across $y = 0$ line. 

We first calculate the current density 
$J_{\perp}$ that is perpendicular to ${\bf k_R}$. Then we find the local 
minimum $J_{\perp, \min}$ along $y = 0$ line. 
The size of the diffusion region is closed by the positions where 
$J_{\perp}$ is half of $J_{\perp, \min}$. The plasma 
flows converge into the left and right diffusion region boundaries 
along $x$-axis are considered as the inflows. The flows diverge out 
from the upper and lower diffusion region boundaries along $y-$axis are 
considered to be the outflows. 
The length of a diffusion region is defined as the distance between 
the upper and lower boundaries along ${\bf k_R}$. 
The inflow characters along the left boundary of the diffusion region 
we choose are influenced by the inflow-outflow coupling. The position of 
this boundary on $y = 0$ line is labeled as the red cross in panel A). 
We average the inflows spatially along a line that centers at this 
point and extends along ${\bf k_R}$ in a length of the diffusion 
region length.

We also measure the magnetic field $B_{\parallel}$ that is 
parallel to ${\bf k_R}$ in the position where the current density 
$J_{\perp}$ is roughly $0$ to the left of the diffusion region inflow 
boundary. As the frozen-in condition is satisfied, the measured magnetic 
field is considered to be the upstream magnetic field of the 
reconnection in this coupled diffusion region. The position of the 
measurement on $y = 0$ line is labeled as the blue cross in panel A).
The magnetic field is averaged along the line that centers at this 
point and extends along ${\bf k_R}$ in a length of the diffusion 
region length.

The normalized spatially averaged inflows 
$\langle V_{\text{in}} \rangle / V_{A0}$ are plotted in panel B) of 
Fig.\ref{fig:0301coupling2}. It can be seen that the local inflow into 
the diffusion region is enhanced by roughly $6$ times on the 
anti-symmetric plane in the double-layer simulation compared to that 
in the single-layer simulation. 
The inflow into the diffusion region on the symmetric plane is 
initially comparable as that in the single-layer simulation. 
It decreases with time and falls to roughly one-quarter of the 
measurement in the single-layer simulation around $t = 50t_A$. 
The result directly shows that the inflow-outflow coupling 
contributes to the local reconnection acceleration. 

The normalized spatially averaged upstream magnetic fields 
$\langle B_{\text{in},\text{e}} / B_0 \rangle$ are plotted in panel C), 
which shows another effect of the inflow-outflow coupling. 
As the inflow increases, the magnetic field is transported fast into 
the diffusion region thus piles up in front of the inflow region. 
It is shown that the upstream magnetic field of the diffusion region 
on the anti-symmetric plane in the double-layer simulation is roughly 
twice of that in the single-layer simulation at $t = 50t_A$. 
In the single-layer simulation, the upstream magnetic field gets 
weaker as the reconnection saturates. Reversely, the upstream magnetic 
field of the diffusion region on the symmetric plane is growing. 
It is probably because that the reconnection in the diffusion region 
saturates so fast that it cannot consume all magnetic fields that are 
transported into it previously. Thus a piling up of magnetic flux is 
observed.

\begin{figure}
   \centering
   \includegraphics[scale=0.28]{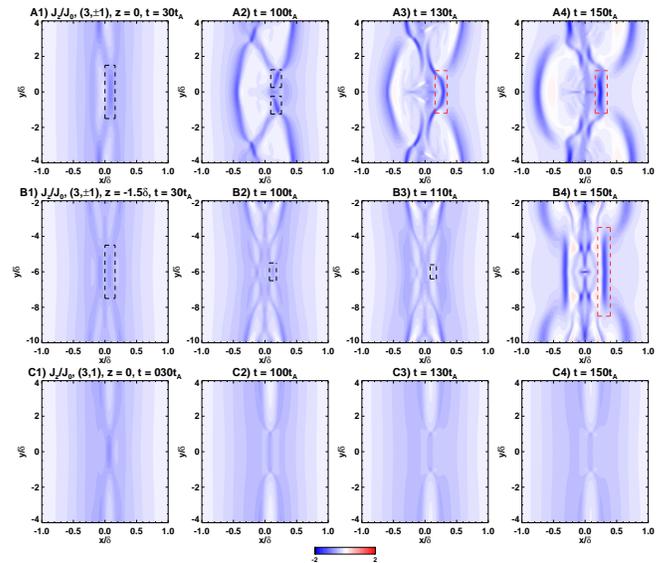}
   \caption{Current density contour plot on anti-symmetric plane $z = 0$ 
            (upper panels) and symmetric plane $z = -1.5\delta$ (middle 
            panels) in the double-layer simulation $(3,\pm 1)$ and 
            $z = 0$ plane in the single-layer simulation (lower panels) 
            along time. 
            The original diffusion regions are labeled by black boxes 
            with dashed lines, while new diffusion regions are labeled 
            by red boxes with dashed lines.
            Only one period along $y$-direction is shown.}
   \label{fig:0301sectearing2}
\end{figure}

Theoretically, secondary tearing is triggered when $\Delta'$ is large 
\citep{jem03,lou05}. It means a strong piling-up magnetic field in 
front of the diffusion layer. Therefore, secondary tearing is expected 
to grow when the inflow-outflow coupling is effective, as the upstream 
magnetic field accumulates. 
In Fig.\ref{fig:0301sectearing2}, we track the diffusion region 
development on the anti-symmetric and symmetric planes in the 
double-layer simulation $(3,\pm 1)$ and on the $z = 0$ plane in the 
single-layer simulation $(3,1)$ along time.
At $t = 100t_A$ in the double-layer simulation $(3,\pm 1)$, 
secondary tearing feature is observed on the anti-symmetric plane in 
A2). No such feature is found on the symmetric plane in the double-layer 
simulation $(3,\pm 1)$ or in the single-layer simulation $(3,1)$.
The diffusion region on the symmetric plane shrinks 
without splitting, as can be observed from B1) to B3). 
The diffusion region in the single-layer simulation extends and 
saturates as the current density becomes weaker from C1) to C4). 
From the information above, it can be seen that the secondary-tearing-like 
feature in $(3,\pm 1)$ is not an intrinsic character of the tearing mode 
$(3,1)$ but rather a result from the interaction of reconnection layers.

Secondary tearing is an indication of a further violent change of 
a current sheet, as shown in the plasmoid instability 
\citep{lou05,hua10,bha09,lou12,shiba15}. 
Unlike the plasmoid instability, which comes from the interaction 
between diffusion regions and plasmoids along a single current sheet, 
the inflow-outflow coupling in 3D is in the direction across the global 
current sheet. The coupling can be only built when there exists a phase 
shift of tearing modes. It reflects the importance of understanding the 
three-dimensionality in the reconnection study.

In the plasmoid instability, diffusion regions are incessantly produced 
during the violent phase. From Fig.\ref{fig:0301sectearing2}, 
it can be found that new diffusion regions are created. 
In addition to the partition of the original diffusion region in A2), 
a new overlying diffusion region starts to grow outside on the 
anti-symmetric plane in the double-layer simulation $(3,\pm 1)$. 
It becomes more visible with time and seems to dominate the reconnection 
at $t = 150t_A$ in A4). 
The production of the overlying diffusion region on the anti-symmetric 
plane is likely to relate with the partitioned diffusion regions. 
As two diffusion regions are produced as A2) shows, their sheet-wise 
outflows have a head-on collision and diverge. The diverging flow 
along positive-$x$ carries the magnetic fields outwardly and meet 
the free field lines that are going inwardly under the global Lorentz 
force. The squeeze then creates the new diffusion region. 

A new overlying diffusion region is also identified outside of the 
original one on the symmetric plane at $t = 150t_A$ in B4). Its 
production mechanism is thought to be different from that on the 
anti-symmetric plane. 
Because no new diffusion region is found in the single-layer simulation 
$(3,1)$, the production should be related to the interaction of 
reconnection layers. 

    \subsubsection{Global development of reconnection layer coupling}\label{sec:result1:typical:global}

\begin{figure}
   \centering
   \includegraphics[scale=0.4]{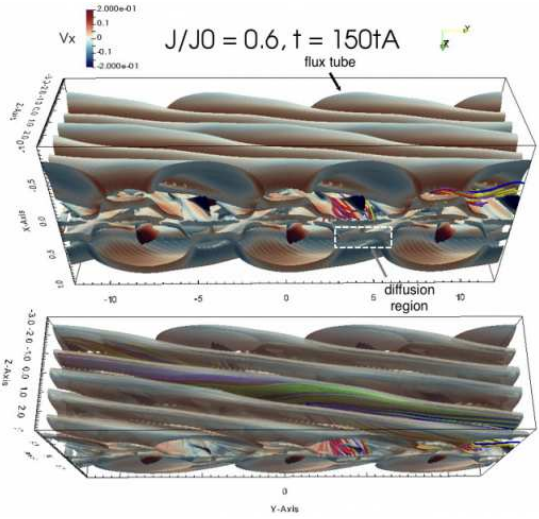}
   \caption{Current density isosurface represents flux tube boundaries 
            and diffusion regions in the whole simulation box. 
            Yellow, purple and green lines are selected magnetic field 
            lines from $z = 0$, $-1.2\delta$ and $1.2\delta$ planes, 
            while red and blue lines are selected magnetic field lines 
            on $z = -0.6\delta$ and $0.6\delta$ planes. 
            All lines root along the layer corresponds to energy mode $(3,-5)$.} 
   \label{fig:0301t150tA2}
\end{figure}

The creation of diffusion regions outwardly to the initial diffusion 
regions on both characteristic planes at $t = 150t_A$ suggests 
that the production is a global trend. 
The current denstiy isosurface $J/J_0 = 0.6$ in the whole simulation box is 
plotted in Fig.\ref{fig:0301t150tA2} to show the global picture. 
The wavy structures are boundaries of flux tubes. There is one group of flux 
tubes on either side of the central global current sheet. They extend along 
the sheet direction. 
The orientation of these flux tubes indicates the growth of new energy 
modes $(3,5)$ and $(3,-5)$, which are growing on different $x-$planes 
compared to the initial energy modes $(3,1)$ and $(3,-1)$. 
By plotting the magnetic field lines that root on the layer corresponds 
to the energy mode $(3,-5)$, it can be seen that these lines are twisting with 
each other and confined inside the flux tube. 
Since these flux tubes cover large areas along global sheet direction, they 
are created by a reconnection layer respectively. 
Therefore, it is proved that new diffusion regions are created globally and 
form a new reconnection layer on either side of the central global current sheet. 

The layers on which energy mode $(3,5)$ and $(3,-5)$ grow are closer to 
the asymptotic magnetic fields than the initial energy modes $(3,1)$ and $(3,-1)$ 
by applying Eq.(\ref{eq:layerpos}). They are capable to transport 
the magnetic fields with large magnetic energy into the global current 
sheet central region to be dissipated. If inflow-outflow couplings still 
exist in this stage to support a fast local reconnection, the global 
reconnection is expected to increase compared to the case when all 
reconnection layers are closer to the global central current sheet center.

\begin{figure}[!t]
   \centering
   \includegraphics[scale=0.55]{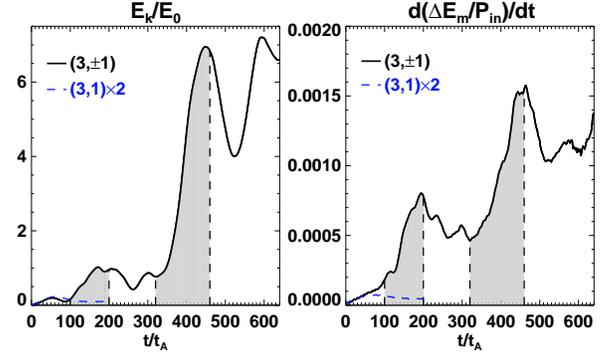}
   \caption{The total kinetic energy (left) and the overall magnetic energy 
            dissipation rate (right) of the double-layer simulation 
            $(3,\pm 1)$. 
            The periods in which the magnetic energy dissipation rate has 
            a boost are shaded by gray. The normalized result from 
            the single-layer simulation $(3,1)$ until $t = 200t_A$ 
            is also plotted as the blue dashed lines.} 
   \label{fig:0301energy}
\end{figure}

We first calculate the total kinetic energy in the box and the overall 
magnetic energy dissipation rate. The total kinetic energy is calculated as:
\begin{equation} \label{eq:Ek}
  E_k = \int_{-\frac{L_z}{2}}^{\frac{L_z}{2}} 
        \int_{-\frac{L_y}{2}}^{\frac{L_y}{2}}
        \int_{-\frac{L_x}{2}}^{\frac{L_x}{2}} 
        \frac{1}{2} \rho {\bf v}^2 dxdydz. 
\end{equation}
The overall magnetic energy dissipation rate is calculated by the time 
derivative of the reduced total magnetic energy: 
\begin{equation} \label{eq:reduceEm}
   \Delta E_m = \int_{-\frac{L_z}{2}}^{\frac{L_z}{2}}
                \int_{-\frac{L_y}{2}}^{\frac{L_y}{2}}
                \int_{-\frac{L_x}{2}}^{\frac{L_x}{2}} 
                     \frac{|{\bf B_f}|^2-|{\bf B}|^2}{8\pi} dxdydz,
\end{equation}
where ${\bf B_f}$ is the magnetic field calculated in a nonlinear 
simulation without any perturbation. By subtracting ${\bf B_f}$, 
the magnetic change from the global diffusion is removed. The overall 
magnetic energy dissipation rate is normalized by 
$P_{\text{in}} = 2L_y L_z v_{A0} B_{y0}^2/(4\pi)$. 
It represents the total inflow Poynting flux estimated by the global
asymptotic magnetic field from both sides of the central global 
current sheet. 

By observing the total kinetic energy and the overall magnetic energy 
dissipation rate in Fig.\ref{fig:0301energy}, we notice two 
boosts in the two shaded periods 
($t = 100 - 200t_A$ and $t = 320 - 460t_A$). In comparison, 
no such eruptive energy release is observed in the single-layer 
simulation $(3,1)$.

The overall magnetic energy dissipation rate boosts suggest 
faster local reconnection in these periods, which can be achieved by 
inflow-outflow couplings from our 
analysis in Sec.\ref{sec:result1:typical:local}. 
In order to build inflow-outflow couplings across the sheet direction, 
multiple reconnection layers that span across the sheet direction 
should coexist in the central global current sheet. 
Thus multiple energy modes with large enough amplitude 
to participate in reconnection on these layers are required.

We calculate the magnetic energy of each mode $(m,n)$ in the Fourier 
space by converting the magnetic field
\begin{equation} \label{eq:Bmode}
  \tilde{{\bf B}}(x,m,n) 
    = \frac{1}{L_y L_z}
      \int_{-\frac{L_z}{2}}^{\frac{L_z}{2}} 
      \int_{-\frac{L_y}{2}}^{\frac{L_y}{2}} 
        {\bf B}(x,y,z) e^{-i(k_y y+k_z z)} dydz
\end{equation}
into magnetic energy spectrum and integrate along the $x$-direction 
that 
\begin{equation} \label{eq:Emmode}
  \tilde{E}_{m}(m,n) 
    = \int_{-\frac{L_x}{2}}^{\frac{L_x}{2}}
        \frac{|\tilde{{\bf B}}(x,m,n)|^2}{8\pi} dx
\end{equation}
to track the dominant energy mode growth along time.
Since we are applying the rotational-symmetric setup, we average all 
energy modes in the form of 
\begin{equation} \label{eq:Emmodeave}
  \tilde{E}_{m}(m,|n|) 
    = \frac{1}{2} 
        \left [ \tilde{E}_{m}(m,+|n|) + \tilde{E}_{m}(m,-|n|) \right ]
\end{equation}
for convenience.

\begin{figure}
   \centering
   \includegraphics[scale=0.55]{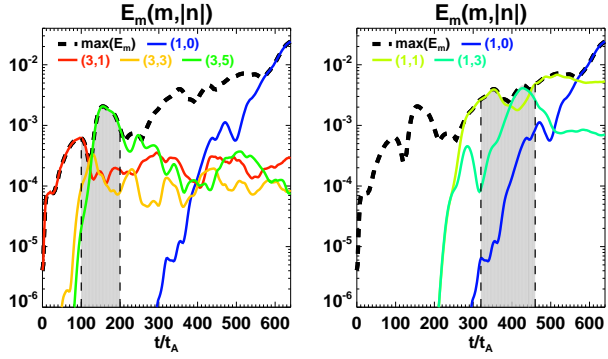}
   \caption{Dominant magnetic mode energy in the selected periods 
            (shaded by gray colors) which correspond to the overall 
            magnetic energy dissipation rate boosts in Fig.\ref{fig:0301energy}. 
            Black dashed line is the maximum mode energy at each time 
            step ($(0,0)$ mode excluded). Colored lines are the most 
            energetic modes during each boost. 
            An overlap of a color line with the black dashed line 
            means that this mode is the dominant mode at this moment.
            The energy mode $(1,0)$ is also plotted as blue lines.} 
   \label{fig:0301spectrum}
\end{figure}

Selected energy modes are plotted in Fig.\ref{fig:0301spectrum}. 
The periods shaded by gray correspond to the same periods in 
Fig.\ref{fig:0301energy}.
A selected mode is the energy mode which grows over other modes, 
namely becomes the most energetic mode, in the gray shaded regions.
The 3D modes (i.e., modes with $n \ne 0$) always dominate during the 
two boosts periods.
Inside the first shaded period ($t = 100 - 200t_A$),
the most energetic mode changes from the initially grown energy 
mode $(3,1)$ to $(3,3)$ then finally $(3,5)$. It shows a shift from an 
energy mode with a smaller $|n|$ to an energy mode with a larger $|n|$. 
Thus it indicates a creation of reconnection layer outwardly. 
On the right panel, we plot the most energetic modes in the second 
shaded period ($t = 320t_A - 460t_A$). 
Before entering this phase, the system seems to undergo a 
re-organization. The energy mode $(1,1)$, which is closer to the 
central global current sheet center than its predecessor $(3,5)$, grows 
rapidly. It corresponds to a coalescence-like process. The energy 
mode $(1,3)$ grows fast after energy mode $(1,1)$. 
It also shows an increase of only $|n|$ value thus new reconnection 
layers grow externally. 
Since the difference of the most energetic modes strength in the 
shaded periods is within 1 order of magnitude, it implies the 
coexistence of multiple reconnection layers in the central global 
current sheet.

None of the successive energy modes in the shaded periods is preferred 
by the initially static system. It means that they must grow by 
receiving energy from the existing energy modes by nonlinear process 
\citep[e.g.,][]{kus87}. We refer to this mechanism as the 
nonlinear coupling of two energy modes $(m_1,n_1)$ and $(m_2,n_2)$ 
in the following context, which can be described as
\begin{equation} \label{eq:4wave}
  (m_1,n_1) + (m_2,n_2) \rightarrow (m_1 \pm m_2,n_1 \pm n_2)
\end{equation}
in general. Chain effect of successive nonlinear coupling 
between daughter and mother energy modes will spread out an 
extended energy spectrum. We calculate the energy transfer rate 
along each path to find the most contributing one that explains 
the emergence of new energy modes in the shaded periods of 
Fig.\ref{fig:0301spectrum}. We skip the second shaded region as 
the whole system becomes so turbulent that the energy mode is 
less well-defined.
The calculation of the energy transfer rate is based on the study 
by \citet{dah92}, which is the extension of the work in hydrodynamics 
\citep{ors83}.

\begin{figure}
   \centering
   \includegraphics[scale=0.55]{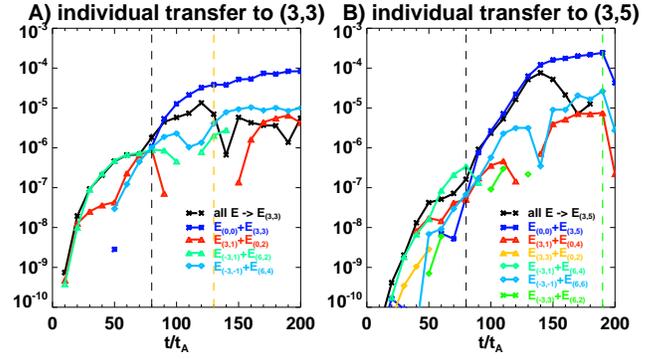}
   \caption{The individual energy transfer rates of energy modes $(3,3)$ 
            and $(3,5)$ from different paths. 
            Black dashed lines ($t = 80t_A$) are the starting time 
            for the rapid growth of energy modes $(3,3)$ and $(3,5)$. 
            Orange ($t = 130t_A$) and green dashed lines ($t = 190t_A$)
            label the peak times for the dissipation rate of 
            these two modes respectively. }
   \label{fig:0301Echain3}
\end{figure}

The individual energy transfer paths of energy modes $(3,3)$ and $(3,5)$ 
are showed in Fig.\ref{fig:0301Echain3}.
Before the rapid growth from $\sim 90t_A$, the main energy transfer 
paths into the energy mode $(3,3)$ are $(3,1)+(0,2)$, $(-3,1)+(6,2)$ and 
$(-3,-1)+(6,4)$. 
The first transfer path can be understood by geometric confinement. 
The second and the third transfer paths both involve higher wavenumber 
modes. Energy mode $(6,2)$ is the second harmonic of the original 
energy mode. Energy mode $(6,4)$ is produced predominantly by $(-3,1)+(9,3)$ 
(not shown) and energy mode $(9,3)$ is the third harmonic of the original mode. 
Similar processes happen for energy mode $(3,5)$. The energy transfer path 
$(-3,1)+(6,4)$ dominates before $(3,5)$ grows rapidly. In addition, the second 
harmonic of energy mode $(3,3)$ is also participating in transferring the 
energy via $(-3,-1)+(6,6)$.

A general trend in both cases is that the new energetic mode in the positive-$x$ 
side is mainly charged by the nonlinear coupling of the present strongest 
energy mode in the negative-$x$ side and higher harmonics (and its product) 
modes in the positive-$x$ side (turquoise lines). 
On the other hand, the growth of higher order harmonics of the original 
energy modes is related to the secondary tearing. 
As the secondary tearing is triggered by the inflow-outflow coupling 
between the initial energetic modes, it is suggested here that the 
inflow-outflow coupling and the nonlinear coupling are related processes.

After $80 t_A$, the energy transfer from energy mode $(0,0)$ becomes 
predominant for both modes. It runs over the total energy transfer from $90t_A$ 
in growing energy mode $(3,3)$ and $140t_A$ in growing energy mode $(3,5)$, 
since these two modes are pouring energy to other modes as well. 
It implies that these two energy modes are growing mainly by 
absorbing energy from the background current after they are charged by 
other modes to a large enough amplitude, similar to an ordinary tearing 
mode.

As new energy modes are activated near the sheet edge due to the nonlinear 
coupling, the reconnection layers span from one side of the central global 
current sheet to the other side. By a direct observation of local flow 
pattern and diffusion region distribution, an increasing number of 
inflow-outflow coupling event can be found inside the central global 
current sheet during the shaded periods. 
Therefore, a 3D web-like energy consuming engine is formed by connecting 
magnetic fields from either side of the sheet, where magnetic energy is 
abundant, through multiple layers in between. 
As the layer extends along the sheet and the local reconnection is fast 
due to the inflow-outflow coupling, a global fast reconnection is achieved 
when new layers are formed near the edges (Fig.\ref{fig:0301energy}).

\begin{figure}
   \centering
   \includegraphics[scale=0.55]{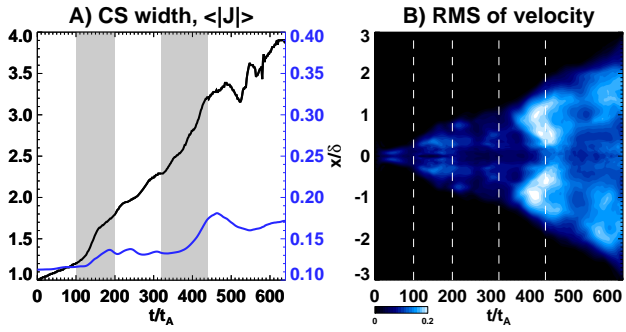}
   \caption{Panel A): the central global current sheet width (black line) and 
            the spatially averaged current density (blue line) in the whole 
            simulation box. 
            Panel B): root-mean-square of the velocity field along the 
            $x$-direction. 
            The two boost periods are shaded by gray in panel A) and in 
            between the dashed lines in panel B).}
   \label{fig:0301CSwidth}
\end{figure}

To consolidate our argument that 3D modes are important in the 
3D reconnection, we plot the spatially averaged current density 
strength $\langle |{\bf J}| \rangle$ 
\begin{equation}
  \langle |{\bf J}| \rangle = \int_{-L_z/2}^{L_z/2} 
                              \int_{-L_y/2}^{L_y/2}
                              \int_{-L_x/2}^{L_x/2}
                               |{\bf J}(x,y,z)|dxdydz/(L_x L_y L_z)
\end{equation}
and central global current sheet width in panel A) of Fig.\ref{fig:0301CSwidth}.
The central global current sheet width is defined as the distance 
between the positions with a certain current density threshold that 
equals to the current density at the initial current sheet half-width
($\sim 0.24J_0$).

Inside the shaded periods which correspond to the reconnection boosts, 
there exist faster expansions of the current sheet. Meanwhile, 
$\langle |{\bf J}| \rangle$ increases during these phases. It 
also indicates an acceleration of global dissipation. During the 
re-organization phase, the current sheet also expands. It is 
mainly due to the coalescence-like process that larger flux tubes 
are created. The expansion speed in the re-organization phase is 
roughly half compared to the fast expansion in the two shaded periods.

In panel B), we plot the root-mean-square of the total velocity 
$v_{\text{rms}}$ along $x$-direction. The main contribution to the 
total velocity is $v_y$, which is the outflow from diffusion regions. 
It can be seen that $v_{\text{rms}}$ increases during two boost periods.
Meanwhile, the enhancement is off from the central global 
current sheet center as they are mainly the outflow from the 3D modes.

From all arguments above, we found how important 3D energy modes are 
for the whole development of the 3D reconnection picture by exploring 
details in one simulation result. In the following part, our arguments 
are rechecked in various conditions to get a general conclusion.

  \subsection{Parameter survey of inflow-outflow coupling}\label{sec:result1:survey}

In this section, we examine the dependence of the local reconnection 
enhancement due to the inflow-outflow coupling on the layer distance. 
In addition, we check the generality of our results on the global development. 
The double-layer simulations used in this section are $(1,\pm 1)$, 
$(2,\pm 1)$, $(3,\pm 1)$ and $(4,\pm 1)$. Their single-layer simulation 
counterparts are also included for comparison.

In our system, changing ${\bf k}$ will change both the reconnection layer 
position $x_s$ and the tearing mode wavelength $\lambda$. Because the inflow 
region size is positively scaled to the wavelength \citep{ste83}, 
we normalize the layer distance $d_l$ to $\lambda$ and define a 
normalized layer distance $A_c$:
\begin{equation} \label{eq:couplingaspect}
  A_c = \frac{d_l}{\lambda} 
      \sim \frac{2\alpha a n}{L_z} 
           \sqrt{1+ \left ( \frac{n L_y}{m L_z} \right )^2} 
\end{equation}
where $d_l$ is evaluated by Eq.(\ref{eq:layerpos}). 
By this definition, the double-layer simulation $(4,\pm 1)$ has the smallest 
$A_c$ value while $(1,\pm 1)$ has the largest $A_c$. 
We check the inflow enhancement dependence on the value of $A_c$ in 
the following part. The diffusion region picked in the analysis is always 
the diffusion region on the layer produced by the tearing mode $(m,1)$ and 
locates across 
$y = z = 0$ line (similar to Sec.\ref{sec:result1:typical:local}). 

\begin{figure}
   \centering
   \includegraphics[scale=0.5]{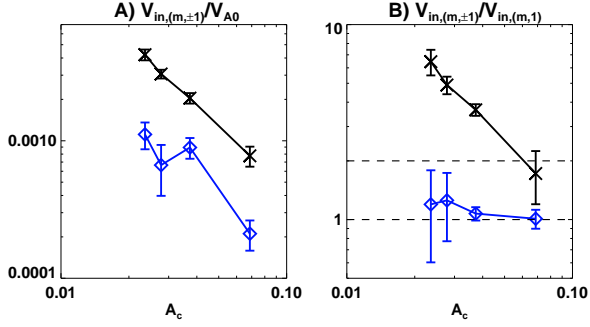}
   \caption{Panel A): time-averaged inflow of diffusion regions on $z = 0$ 
            plane in double-layer simulations. 
            Panel B): time-averaged inflow ratio of double-layer simulations 
            to their correspondent single-layer simulations. 
            Black lines with crosses are the measurements on the left of 
            the diffusion region. Blue lines with diamonds are the measurements 
            on the right of the diffusion region. 
            Standard deviations are shown as error bars.}
   \label{fig:mninflowplot}
\end{figure}

We measure the inflow strength on both sides of the diffusion region on 
$z = 0$ plane in all simulations. The inflow measuring position is similar 
to that in Sec.\ref{sec:result1:typical:local}.
The measurements are averaged along ${\bf k_R}$ in a length half of the 
diffusion region length, then averaged along time from when the flow 
becomes steady until secondary tearing starts to grow.
The inflow from the left side of the diffusion region is the inflow 
involved in the inflow-outflow coupling in a double-layer simulation. 
The inflow from the right side of the diffusion region does not 
participate in the inflow-outflow coupling.
The result is shown in Fig.\ref{fig:mninflowplot}.

The inflow measured from the left side increases monotonically in panel A) 
when the reconnection layers are closer to each other. In order to see 
how much the inflow from the left side enhances due to the inflow-outflow 
coupling, we normalize the inflow in the double-layer simulations to their 
single-layer counterparts and plot it in panel B). 
The inflow enhancement also shows a monotonic increase with decreasing 
layer distance. It is worth pointing out that the inflow ratio for 
$(1,\pm 1)$ with $A_c \sim 0.07$ is $\sim 2$. It means the 
inflow-outflow coupling is ineffective that the resultant inflow 
is roughly the superposition of the inflow into the present detecting 
diffusion region and the diverging outflow from the diffusion region 
on the other side of the current sheet. 
With a smaller distance between layers ($m \ge 2$), an effective 
inflow-outflow coupling is built that the enhancement value 
exceeds $2$ largely. 

In comparison, it is shown in panel B) that the inflow in double-layer 
simulations from the right side is not effectively changed compared 
to single-layer simulations even when the inflow-outflow coupling is 
strong from the left side. Nonetheless, the entire diffusion region shows 
an increased potential in reconnection when the layers are closer to 
each other.

\begin{figure}
   \centering
   \includegraphics[scale=0.5]{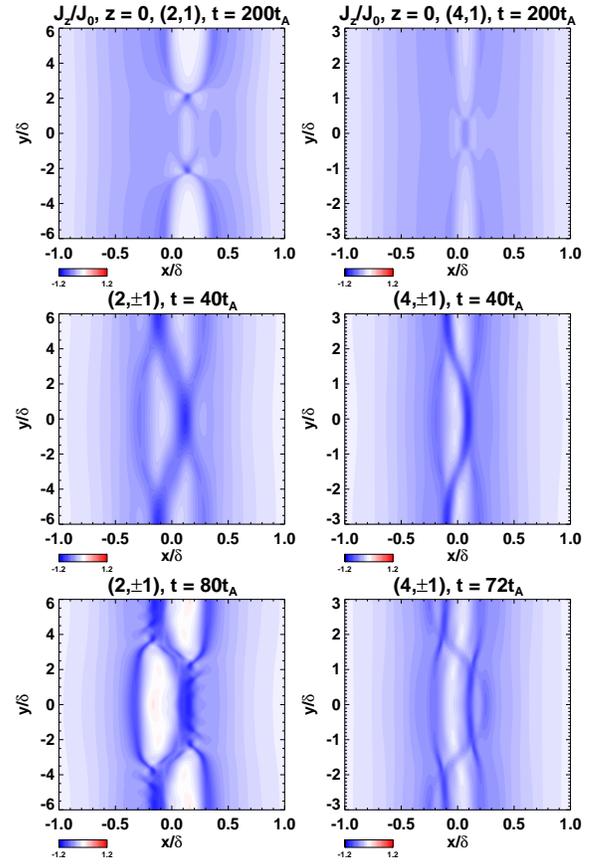}
   \caption{Current density $J_z$ contour plot for double-layer simulations 
            $(2,\pm 1)$, $(4,\pm 1)$ and their correspondent single-layer 
            simulations at different times on the anti-symmetric plane $z = 0$.} 
   \label{fig:mnJzasym}
\end{figure}

From our result in Sec.\ref{sec:result1:typical:local}, 
we have found that the inflow-outflow 
coupling will trigger secondary tearing even if the original single 
tearing mode is stable to secondary tearing. In Fig.\ref{fig:mnJzasym}, 
we compare the double-layer simulations $(2,\pm 1)$, $(4,\pm 1)$ and 
their correspondent single-layer simulations to see the local 
development of diffusion regions in the early stage. 
In the upper panels, it can be seen that the tearing mode $(2,1)$ is 
potentially vulnerable to the secondary tearing, while mode $(4,1)$ 
is secondary-tearing-stable. Both double-layer simulations $(2,\pm 1)$ 
and $(4,\pm 1)$ exhibit the character of secondary tearing that 
partition of current density peak can be seen from the lower panels. 

\begin{figure}
   \centering
   \includegraphics[scale=0.55]{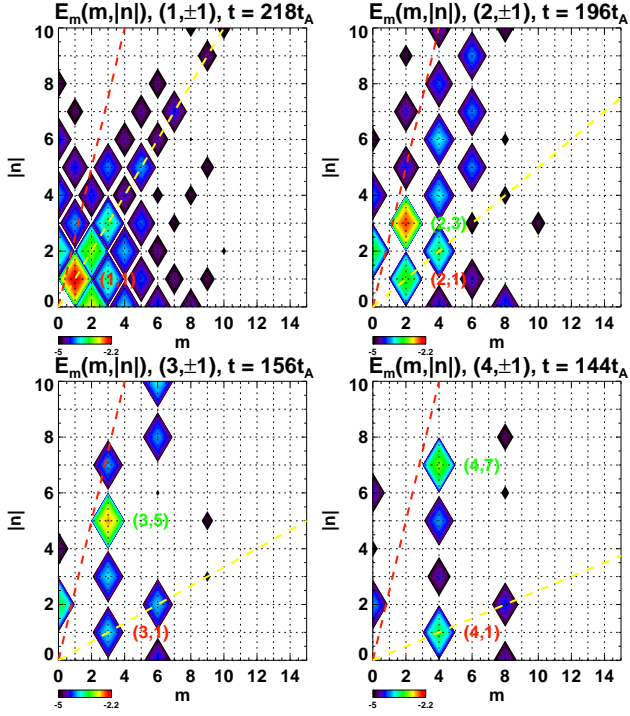}
   \caption{Magnetic energy of energy modes calculated by 
            Eq.(\ref{eq:Emmodeave}) for all double-layer simulations. 
            The original energy modes are labeled as red characters, 
            while the green color represents the final energy mode in 
            the energy transfer. 
            The red dashed line is the asymptotic magnetic field 
            orientation. The yellow dashed line is the 2D-like cascade 
            path for initially perturbed modes. 
            Colors are in the logarithmic scale.}
   \label{fig:all2Dspec}
\end{figure}

The magnetic energy of energy modes calculated by Eq.(\ref{eq:Emmodeave}) 
at a certain time for all double-layer simulations is plotted in 
Fig.\ref{fig:all2Dspec}. 
The time slice for the double-layer simulation $(1,\pm 1)$ is chosen when 
the overall magnetic energy dissipation rate reaches the first peak. 
The others are chosen when the final energy modes in the energy transfer 
reach their peaks. 
In the effective inflow-outflow coupling cases ($m \ge 2$), a energy transfer 
path along the global guide field ($k_z$-direction, or along $|n|$) 
can be seen. However, in the double-layer simulation $(1,\pm 1)$, 
the energy mainly goes along the 2D-like cascade path (yellow dashed line). 

\begin{figure}
   \centering
   \includegraphics[scale=0.55]{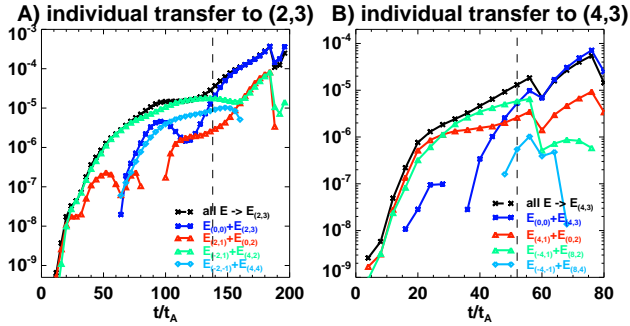}
   \caption{Individual energy transfer rate of energy modes $(2,3)$ and 
            $(4,3)$ in double-layer simulations $(2,\pm 1)$ and $(4,\pm 1)$, 
            respectively.
            Black dashed lines at $t = 138t_A$ in A) and $t = 52t_A$ in B) are the 
            starting time for the rapid growth of $(2,3)$ and $(4,3)$.}
   \label{fig:2141Echain2}
\end{figure}

We explore the individual energy transfer paths for both double-layer 
simulations $(2,\pm 1)$ and $(4,\pm 1)$. 
The energy transfer rate of energy modes $(2,3)$ and $(4,3)$, which 
are the first energy mode along $|n|$ direction in the energy transfer, 
is plotted in Fig.\ref{fig:2141Echain2}. 
The turquoise lines of $(-2,1)+(4,2) \rightarrow (2,3)$ in panel A) 
and $(-4,1)+(8,2) \rightarrow (4,3)$ in panel B) are similar paths 
to $(-3,1)+(6,2) \rightarrow (3,3)$ in Fig.\ref{fig:0301Echain3} 
panel C). All of them are dominant before the 
daughter modes are capable to grow themselves by extracting the 
energy from the background magnetic field. 

\begin{figure}
   \centering
   \includegraphics[scale=0.5]{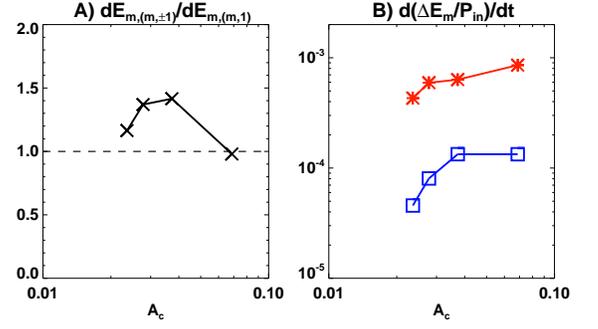}
   \caption{Panel A): the overall magnetic energy dissipation rate ratio of 
            double-layer simulations over their correspondent single-layer 
            simulations just before secondary tearing emerges. The values 
            are normalized by the initial overall magnetic energy 
            dissipation rate ($t \sim 0$). 
            Panel B): the overall magnetic energy dissipation rate of 
            double-layer simulations. 
            The blue line with squares is the measurements just before 
            secondary tearing emerges, the same as panel A). 
            The red line with stars is the measurements selected at 
            the same times as that in Fig.\ref{fig:all2Dspec}.}
   \label{fig:mnMRplot}
\end{figure}

Finally, we compare the overall magnetic energy dissipation rate 
calculated by Eq.(\ref{eq:reduceEm}) of all simulations 
(Fig.\ref{fig:mnMRplot}). We measure at two different times. 
The first measurement is taken just before secondary tearing emerges. 
The second measurement is taken at the times the same as that in 
Fig.\ref{fig:all2Dspec}.

In the understanding of the first measurement, we compare the overall 
magnetic energy dissipation rate of double-layer simulations over their 
correspondent single-layer simulations in panel A). 
Although the inflow-outflow coupling becomes more effective with 
decreasing $A_c$, the overall magnetic energy dissipation ratio exhibits a 
non-monotonic change with a peak around $A_c \sim 0.035$. 
Probably because the closest pair of the coupling has the 
shortest wavelength, they influence only a limited region near 
the central global sheet center. 
Therefore, the overall magnetic energy dissipation 
rate does not change much compared to its single-layer counterpart. 
The absolute values of the overall magnetic energy dissipation rate 
of all double-layer simulations measured at two different times 
are shown in panel B). 
The double-layer simulation $(1,\pm 1)$ has the largest overall 
magnetic energy dissipation rate values than all other simulations, 
regardless of the time. Nonetheless, the overall magnetic energy 
dissipation rate in the second measurement shows that more effective 
inflow-outflow coupling will result in a higher enhancement of 
the overall magnetic energy dissipation rate compared to their first 
stage. 

From our results above, we have found that the inflow-outflow 
coupling can be built when the relative distance between 
two reconnection layers is small. It can trigger secondary tearing, 
energy transfer to energy modes that grow outwardly and a faster overall 
magnetic energy dissipation rate compared to a single reconnection layer. 
The arguments are retested in the random-perturbation 
simulation group, which has a more general setup and different 
parameter ranges.

\section{Simulation result II: Random-perturbation group}\label{sec:result2}

The complicated structure in the random-perturbation group 
makes the quantitative analysis difficult, especially the local 
analysis as that in Sec.\ref{sec:result1:typical:local}. 
Therefore, we only concentrate on the global character of this 
group.

\begin{figure}
   \centering
   \includegraphics[scale=0.5]{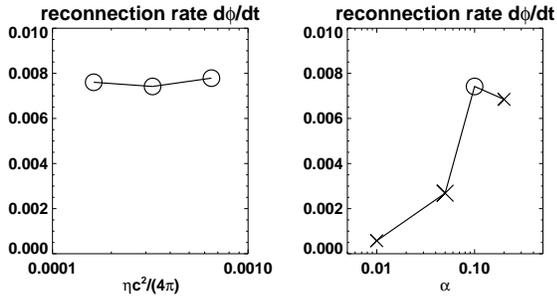}
   \caption{Reconnection rate calculated by the fastest reconnected 
            flux growth rate for each subgroup.}
   \label{fig:randomMRrate2}
\end{figure}

We calculate the reconnection rate by measuring the reconnected 
flux growth rate, which follows the same method as that in 
\citet{hua16}. The fastest reconnected flux growth rate before the 
boundaries start to interfere is shown in Fig.\ref{fig:randomMRrate2}. 
The reconnection rate in $\alpha = 0.01$ is calculated at 
$t = 1000t_A$.
In addition, we calculate the magnetic energy of energy modes. 
The simulations in which the energy transfer along the 
global guide field is found are labeled with circles. Otherwise, 
the simulations are labeled with crosses.

From our simulation result, the reconnection rate of the simulation 
subgroup with different diffusivity ($\alpha = 0.1$) shows a weak 
dependence on the uniform diffusivity. 
The values are in the same order of magnitude as other spontaneous 
3D reconnection studies with moderate global guide field strength 
\citep{kow17,ber17}. All simulations in this subgroup exhibit an 
energy transfer along the global guide field. 
A qualitative examination shows that they basically 
follow the same scenario as what we have found in the 
eigenmode-perturbation group. However, unlike that in the 
eigenmode-perturbation group, multiple reconnection layers (more than 2) 
emerge in the initial state due to the implementation of a random velocity 
field. The distance between layers on the same side of the central 
global current sheet is closer than the distance between layers 
from either side of the sheet. Therefore, the layers interact 
with each other on either side of the central global current sheet 
first. When they grow thick enough, they interact across the 
sheet center and achieve fast global reconnection.

In comparison, the reconnection rate shows a non-monotonical 
dependence on the uniform global guide field. 
It seems that there is a preferred window selected by the global 
guide field for fast energy dissipation in our system. 
When the guide field is small that $\alpha < 0.1$, the global 
reconnection rate reduces with the global guide field 
strength.

In the case with $\alpha = 0.01$, reconnection happens only 
near the central global current sheet center before $t = 1000t_A$. 
Energy modes $(1,1)$, $(1,-1)$ and $(1,0)$ are dominant before 
$t \sim 500t_A$. 
Due to the resonance condition of the closed box, 3D energy modes 
$(1,1)$ and $(1,-1)$ are growing close to the central global current 
sheet center. No energy transfer along the global guide field is 
found. Thus no new reconnection layer grow further away from the 
center. As we have shown in Sec.\ref{sec:result1:survey}, 
the global reconnection rate is low if the reconnection layers are 
confined near the sheet center. On the other hand, 
from $t \sim 500t_A$ until the end of the simulation ($t = 1000t_A$), 
the dominant energy modes are 3D energy modes with $m = 0$. 
These modes do not participate in the reconnection 
thus the global reconnection decreases from $t \sim 600t_A$. 
The reconnection rate increases again from $t \sim 900t_A$ due to 
the growth of 2D energy mode $(1,0)$. 
There remains a possibility that the reconnection can accelerate 
further after $1000t_A$. Since the growth rate of energy mode $(1,0)$ 
is $1/4$ of that in the random perturbation simulation with $\alpha = 0.1$ 
in the similar stage, we believe that the reconnection rate enhancement 
might not be considerable.

In the simulations $\alpha = 0.05$ and $\alpha = 0.2$, the distance 
between initially emerged reconnection layers is larger than the 
distance between reconnection layers in $\alpha = 0.1$ case. 
It is possible that the diffusion regions on the layers in 
$\alpha = 0.05$ and $\alpha = 0.2$ hardly couple so no energy 
transfer along the global guide field is achieved. 
Therefore, the global reconnection is expected to be slow 
following our scenario.

However, $\alpha = 0.2$ case reaches a global reconnection rate 
that is similar to $\alpha = 0.1$ cases. This is probably related 
to the closed boundary condition. 
In our closed system, the global magnetic field structure ends 
with the structure that corresponds to the energy modes with the 
longest wavelengths that can be reached in the system, namely 2D 
energy mode $(1,0)$, 3D energy modes $(1,1)$ and $(1,-1)$.
By Eq.(\ref{eq:layerpos}), the $x$-positions of energy modes $(1,1)$ 
and $(1,-1)$ in $\alpha = 0.2$ case are the largest among all 
simulation cases. 
Since the energy modes $(1,1)$ and $(1,-1)$ are close to the 
asymptotic magnetic field, they can efficiently convert the magnetic 
energy.

\section{Discussion}\label{sec:discussion}

One global picture of the spontaneous 3D reconnection was given in
\citet{ber17}. This work proposed that turbulence gradually builds 
and expands the whole current sheet, finally produces global 2D-like
bi-directional inflows and outflows. 
In our study, we showed how new energy modes, which grow new 
reconnection layers, are produced outwardly. This process explains 
the global current sheet expansion stage. 
Meanwhile, the nonlinear coupling of the pre-existing 
modes creates new modes and builds the turbulent state, as more and 
more energy modes are activated seen from 2D energy spectrum in
Sec.\ref{sec:result1:typical:global}. In the turbulence reconnection 
picture \citep{laz99}, the field lines are thought to wander around 
inside the whole current sheet and gradually release the energy by 
continuous reconnection at different positions. 
We describe this wandering motion as the inflow-outflow coupling of 
diffusion regions on adjacent reconnection layers. The physics of the 
inflow-outflow coupling was examined locally in detail. 
In this sense, we have given a clear physical picture of the 
turbulence reconnection process, from the viewpoint of the diffusion 
region development and the interaction between reconnection layers. 

All our present simulation results are based on the simulation 
in a closed box. It is often the case that the system is open on both 
ends or at least one. 
When the system is open, the reconnected structures can be removed 
from the current sheet by the global outflow. 
From our result in a closed box, the growth of new energy modes away 
from the current sheet center is important in enhancing the global 
reconnection efficiency. It requires $\sim 100t_{A}$ to grow a 
new energy mode by observing Fig.\ref{fig:0301spectrum} in our 
present simulation setup. If we assume the 2D outflow is in the 
magnitude of $v_{A0}$, then the global current sheet must be longer 
than $100\delta$ to ensure that the new energy modes can be 
generated with the present structure. Whether our scenario 
can be retrieved in an open system needs to be examined in a 
simulation box with open boundaries.

From the simulation result of the global guide field subgroup, 
it is found that the coexistence of multiple reconnection layers 
in a closed box does not always lead to a fast global reconnection. 
This is due to the large distance between initially emerged 
reconnection layers, as some tearing modes are suppressed by the 
closed boundary condition. When the simulation box becomes open, 
the tearing modes are free to grow. 
If the distance between the layers on which tearing modes grow is 
small enough to build the inflow-outflow coupling, the global 
current sheet might follow our scenario to reach a fast overall 
reconnection. However, if the distance between layers is in the 
same magnitude as the thickness of the diffusion layer of a 
tearing mode, this bunch of reconnection layers possibly behaves 
like a thick layer. 
How the reconnection develops in such a case needs further 
investigation.

Previous simulation results suggest that the physics in the 
reconnecting global current sheet changes with the applied global 
guide field strength \citep{ois15,kow17}.
We believe that our scenario is applicable directly to describe the 
3D reconnection in a global current sheet with a moderate global 
guide field ($\alpha > 0$). 
Meanwhile, since only one reconnection layer at the 
global current sheet center is allowed to grow when $\alpha \sim 0$, 
the depletion of reconnection layers on which 3D energy modes grow 
implies a slower reconnection by our analysis. 
Therefore, we are capable to give a consistent theory to explain 
the reconnection rate difference in current sheets with 
different global guide field strengths. 
On the other hand, how the turbulence accelerates the 3D reconnection 
in a global current sheet with $\alpha = 0$ is still not fully 
understood. It was thought that Kelvin–Helmholz-like 3D instability 
which breaks the Sweet-Parker \citep{swe58,par57} type diffusion region 
into filaments is important in enhancing the global reconnection rate 
\citep{ois15}. 
Quantitative analysis is needed to have a concrete conclusion.

The diffusivities applied in our simulation are much higher than 
the astronomical value. From our present diffusivity subgroup 
result, a fast reconnection is expected if the diffusivity is 
even lower. However, the developing time to reach a fast 
reconnection might be diffusivity-dependent. 
Assume that the inflow-outflow coupling between diffusion regions 
is a universal phenomenon. 
The tearing mode needs time to grow into an amplitude large 
enough to start the inflow-outflow coupling effectively across 
the global current sheet. The linear growth rate of the tearing 
mode is reversely scaled to the diffusivity \citep{FKR63,cop76}. 
Then it is highly possible that it takes a long time for a system 
to develop the inflow-outflow coupling when the diffusivity is 
small. Once it is built, the inflow-outflow coupling changes the 
spontaneous reconnection into the driven regime locally 
\citep{pri80}. Theoretically, the local reconnection is less 
dependent on the diffusivity in the driven reconnection but rather 
relies on the inflow 
electric field ($E_{\text{in}} \sim v_{\text{in}} B_{\text{in}}/c$) 
\citep{sat79,kit96}. Thus it is possible that the global reconnection 
becomes less dependent on the diffusivity due to the local driven 
reconnection that spreads all over the global current sheet. 
Whether this holds true in a slowly diffusing magnetic field is 
unclear and requires studies set with much lower diffusivity.

\section{Conclusion}\label{sec:conclusion}

In our present study, we found a growth of new energy modes near the 
global current sheet boundaries in spontaneous 3D reconnection. 
The emergence of the new modes comes from the interaction of 
reconnection layers closer to the global current sheet center.  
The global fast magnetic energy dissipation is achieved 
when reconnection layers expand across the current sheet, while 
individual diffusion regions on different reconnection layers 
build an inflow-outflow coupling that supports a fast local 
reconnection.
As a whole, we present a detailed explanation of the 
development of diffusion regions in a current sheet when the 
turbulence is building. We also show how important 3D energy 
modes contribute to the overall energy dissipation. 

Although our simulations are executed in a closed box, there remains a 
possibility that similar processes could be achieved in an open system. 
Further examination with a larger box or with open boundaries should be 
done to confirm our picture.
On the other hand, our present diffusivity is still far higher than 
the astronomical value. Lundquist number larger than $10^7$ 
is required to observe a sufficient energy spectrum.  
Huge simulation on the topic of reconnection is always required to 
explain the enormous energy release in the astronomical phenomenon.

\begin{acknowledgments}
Numerical computations were [in part] carried out on Cray XC30 at 
Center for Computational Astrophysics, National Astronomical 
Observatory of Japan.
Numerical computations were [in part] carried out on Cray XC50 at 
Center for Computational Astrophysics, National Astronomical 
Observatory of Japan.
Takaaki Yokoyama is supported by JSPS KAKENHI Grant: JP15H03640.
\end{acknowledgments}

\nocite{*}
\bibliography{WangMR}

\end{document}